\documentclass[a4paper]{llncs} 

\usepackage{graphicx}
\usepackage[usenames,dvipsnames,svgnames,table]{xcolor}
\usepackage[utf8]{inputenc}
\usepackage{soul}

\usepackage[english]{babel}
\usepackage{amsmath}
\usepackage{amsfonts}
\usepackage{amssymb}
\usepackage{array}
\usepackage{textcomp}
\usepackage{czt}

\usepackage{circus}

\usepackage{hijac}

\usepackage{multirow}

\usepackage{hyperref}
\hypersetup{
    colorlinks,
    citecolor=black,
    filecolor=black,
    linkcolor=black,
    urlcolor=black
}

\usepackage{listings}
\author{Matt Luckcuck \and Ana Cavalcanti \and Andy Wellings}

\institute{Department of Computer Science, University of York, York, YO10 5GH, UK}

\title{A Formal Model of the \\Safety-Critical Java Level 2 Paradigm}

\setul{0ex}{}
\newcommand{\TightRope}{T\textsuperscript{\ul{ight}}R\textsuperscript{\ul{ope}}} 

\begin{document}

\maketitle

\begin{abstract}

Safety-Critical Java (SCJ) introduces a new programming paradigm for applications that must be certified. The SCJ specification~(JSR 302) is an Open Group Standard, but it does not include verification techniques. Previous work has addressed verification for SCJ Level~1 programs. We support the much more complex SCJ Level~2 programs, which allows the programming of highly concurrent multi-processor applications with Java threads, and wait and notify mechanisms. We present a formal model of SCJ Level~2 that captures the state and behaviour of both SCJ programs and the SCJ API. This is the first formal semantics of the SCJ Level~2 paradigm and is an essential ingredient in the development of refinement-based reasoning techniques for SCJ Level~2 programs. We show how our models can be used to prove properties of the SCJ API and applications.  

\end{abstract}
 
\section{Introduction}

Safety-Critical Java (SCJ)~\cite{SCJ_Spec_0.100} is a version of Java that embeds a new programming paradigm for applications that must be certified for example, using the highest level of the avionics standard ED-12/DO-178~\cite{ED12C_DO178C}. To aid certification, SCJ is organised into three compliance levels. Level~0 applications are simple single-processor programs executed by a cyclic executive. Level~1 applications introduce concurrency and less-restricted release patterns. By contrast, Level~2 applications are highly concurrent, potentially multi-processor, and make use of suspension and a variety of release patterns.

The verification of SCJ programs requires specific techniques, but these are not covered by the SCJ specification. Verification has been addressed for Level~1, but not Level~2. SCJ, and its Level~2 profile in particular, present several challenges for verification. The new programming paradigm of SCJ restricts the program structure and provides a predictable memory model. The unique features of Level~2 allow programming applications that may contain multiple modes of operation or independently developed subsystems, and computations that require non-standard release patterns or suspension~\cite{wellings_luckcuck_cavalcanti_2013}.

In this paper, we provide support for verification of SCJ Level~2 programs by modelling its programming paradigm using the state-rich process algebra \Circus{}~\cite{circus_woodcock_cavalcanti_2002}. This is a combination of Z~\cite{spivey1992z} for modelling state, CSP~\cite{hoare1985csp} for modelling behaviour, and Morgan's refinement calculus~\cite{morgan1990programming}. A \Circus{} program is organised around processes, which contain variables and actions, to describe a data model and reactive behaviours. Each process has a main action that defines its behaviour, possibly using a combination of other actions in the process. Communication between processes is achieved via channels. In our work we use the \Circus{} extensions \textsf{\textit{Oh}}\Circus{}~\cite{cavalcanti_sampaio_woodcock_2005}, which introduces object orientation and inheritance, and \Circus{} Time~\cite{Sherif2009-de} to specify timers and deadlines.  

\begin{figure}[t]
\begin{center}
\includegraphics[width=\textwidth]{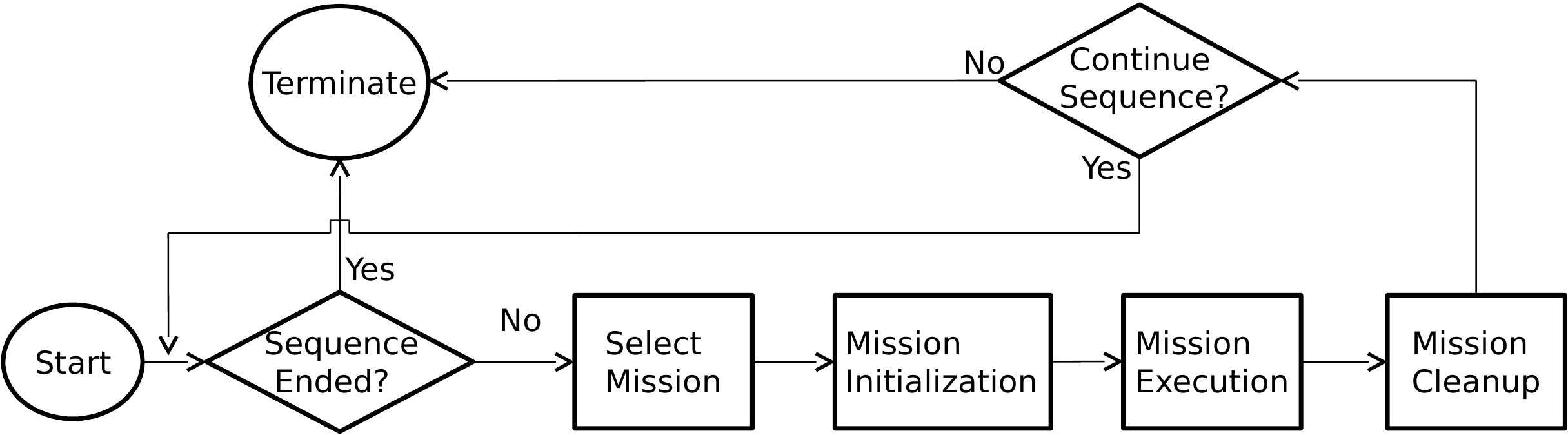}
\caption{Mission Phases \label{fig:phases}}
\end{center}
\end{figure}

\Circus{} has already been used to model SCJ Level~1~\cite{zeyda_etal_2013}. \Circus{} has also been used to produce a refinement strategy~\cite{scjincircus_cavalcanti_2011} to derive SCJ programs that are correct by construction. Our models provide the the possibility of extending the refinement strategy to target SCJ Level~2 programs. 
 
What we present in this paper is the first formalisation of SCJ Level~2. The SCJ API covers approximatively 112 pages of the specification~\cite{SCJ_Spec_0.100} as a collection of approximately 36 classes and interfaces. Our work characterises a semantics for SCJ Level~ 2 programs. To support its use, we have developed a tool that generates \Circus{} models from SCJ programs. We have used the models to prove, via model checking, properties of both the SCJ API and of specific programs.  

In Sect.~\ref{sec:scjl2} we describe the unique features of the SCJ Level~2 paradigm. Section~\ref{sec:modelStructure} describes our modelling approach, model structure, and how we model Java synchronisation and suspension behaviour. Section~\ref{sec:applications} describes the direct applications of our models for verification, including a brief account of our tool. Section~\ref{sec:related} presents related work. Finally, Sect.~\ref{sec:conclusion} concludes this paper with a summary of our contribution and a discussion of future work.

\section{Safety-Critical Java Level 2 Paradigm}
\label{sec:scjl2}

Safety-Critical Java (SCJ) is a version of Java that adopts a new programming paradigm. SCJ programs have a specific concurrent design and use region-based memory management (instead of garbage collection); specialised virtual machines~\cite{Schoeberl2008265,Sondergaard2012-cq} are available to execute SCJ programs. SCJ also uses the real-time constructs introduced in the Real-Time Specification for Java~\cite{RTSJ}, but enforces a more structured programming paradigm. 

An SCJ program is controlled by a \emph{safelet} object, which manages the top-level \emph{mission sequencer}. This is used to activate an application-defined sequence of \emph{missions}. A mission encapsulates a particular function or phase of operation as a set of \emph{schedulable objects} to perform a particular task. An SCJ API supports the programming of these components.

Each mission progresses through an initialisation, execution, and cleanup phase, as shown in Fig.~\ref{fig:phases}. During initialisation, a mission's schedulable objects are created and registered. These schedulables are activated simultaneously at the start of the execution phase. A mission's schedulables execute until one of them requests termination, or they all terminate, when a clean\-up phase is performed. At the end of the clean\-up phase, the mission may indicate that no further missions should execute, in which case the sequence will terminate. If not, and there are more missions to run, the next mission is prepared.

At Level~2, schedulable objects may adopt one of four release patterns. Periodic event handlers execute once in a given time period, aperiodic event handlers execute when triggered by a method call, one-shot event handlers execute once after a time offset, and managed threads simply run to completion. Level~2 supports the execution of concurrent missions by allowing missions to manage schedulable mission sequencers. Level~2 can also use Java suspension methods, \texttt{wait()} and \texttt{notify()}, but they may only be called on \texttt{this}.

\label{sec:exampleApp}
To illustrate some of the features of SCJ Level~2 programs we introduce FlatBuffer, which is a simple solution to the Producer-Consumer Problem, using a one-place buffer. FlatBuffer is structurally simple, only containing one mission and two schedulables, but uses two of Level~2's unique features: managed threads and suspension. Larger examples of applications that use the unique features of Level~2 can be found in~\cite{wellings_luckcuck_cavalcanti_2013}.

\begin{figure}[t]
\centering
\includegraphics[scale=0.5]{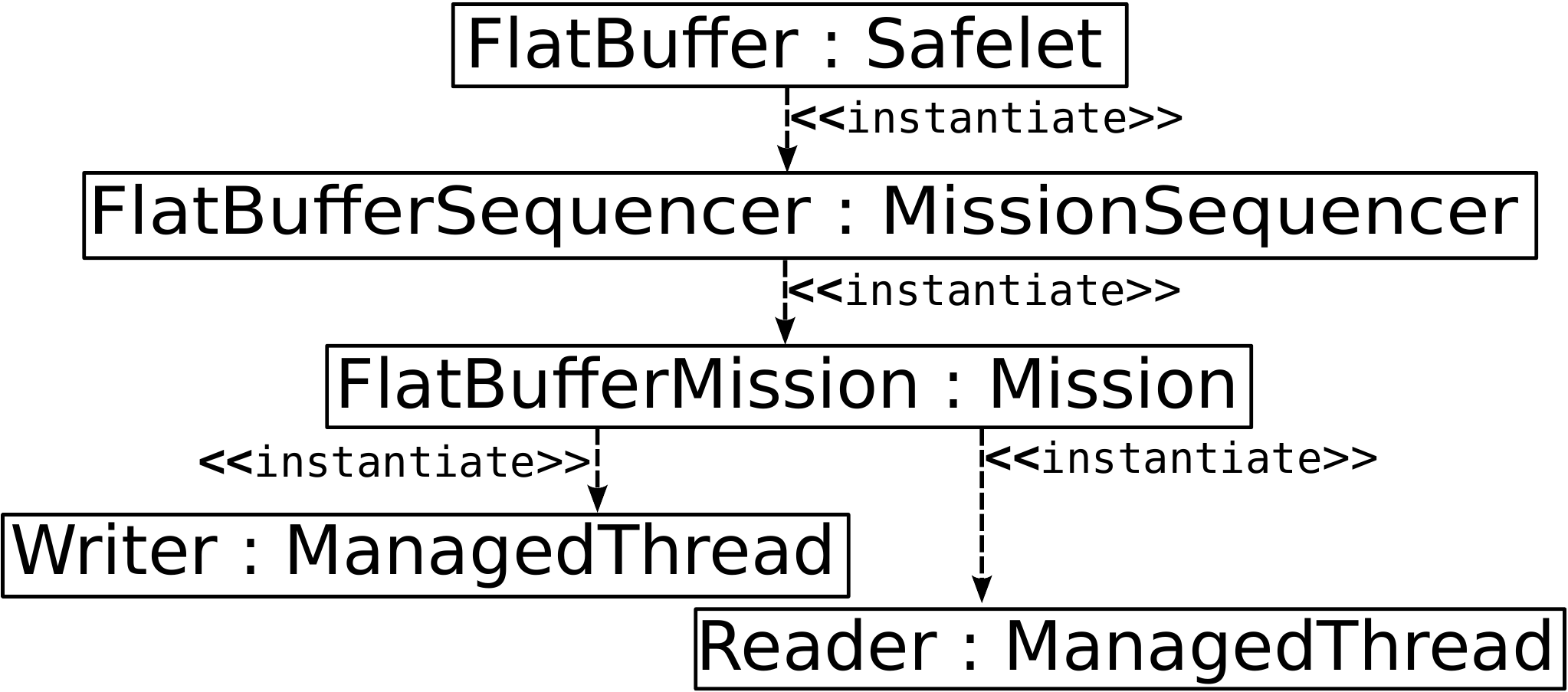}
\caption{Object Diagram of the Flatbuffer \label{fig:flatbuffer}}
\end{figure}

Figure~\ref{fig:flatbuffer} shows an object diagram of the FlatBuffer program at the end of its mission's initialise phase. It is controlled by the safelet \texttt{FlatBuffer}, which starts the top-level mission sequencer \texttt{FlatBufferMissionSequencer}. This mission sequencer starts the mission, \texttt{FlatBufferMission}, which starts the two managed threads. The \texttt{Writer} is the producer and the \texttt{Reader} is the consumer. 

The \texttt{FlatBufferMission} holds the buffer and controls access to it. The mission has a \texttt{bufferEmpty()} method to indicate if the buffer if empty or full, a \texttt{read()} method to control reading from and resetting the buffer, and a \texttt{write()} method to control updating the buffer. The \texttt{read()} and \texttt{write()} methods both use synchronisation to control access to the buffer.

\begin{figure}[t]
\centering
\includegraphics[scale=0.5]{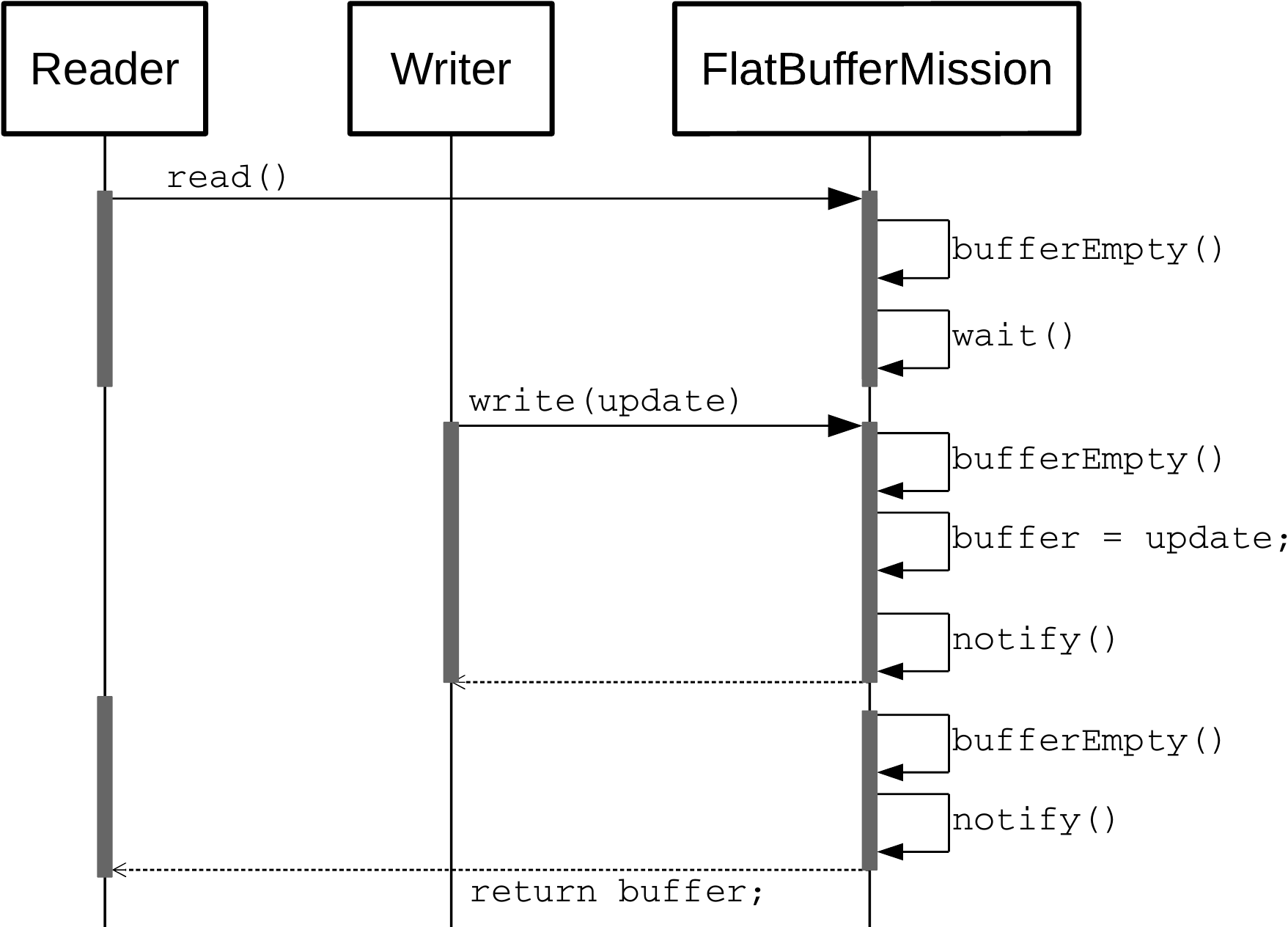}
\caption{Sequence Diagram of an Example Execution of FlatBuffer \label{fig:seqDia} }
\end{figure}

In an example execution of FlatBuffer, illustrated in Fig.~\ref{fig:seqDia}, the \texttt{Reader} runs first, and calls the mission's \texttt{read()} method. The method calls \texttt{bufferEmpty()} on the mission, which returns a boolean indicating that the buffer is empty. Because there is nothing to read, the method calls \texttt{wait()} to suspend the \texttt{Reader}. 

Next, the \texttt{Writer} runs, calling the mission's \texttt{write()} method. This method calls \texttt{bufferEmpty()} on the mission, which still indicates that the buffer is empty, prompting the \texttt{Writer} to update the buffer. Then, the method calls \texttt{notify()} on the mission -- which resumes the \texttt{Reader}. When the \texttt{Reader} resumes, it is still inside the \texttt{read()} method. The method calls \texttt{bufferEmpty()}, which indicates that the buffer is full, so the value is read and the buffer is reset. Since this is a simple test program, the \texttt{Writer} terminates the mission after 5 writes. 

Despite SCJ's restricted infrastructure, the unique features of Level~2 mean that its programs can become very complex. Providing the first semantics for this paradigm and devising a model for Level~2 programs is, therefore, a challenging task. We need to deal with a variety of schedulable objects, a preemptive scheduler that guarantees absence of priority inversion, a complex protocol for termination of missions, and suspension in the context of all of these features. We discuss our approach to modelling SCJ Level~2 in the next section.

\section{Modelling Approach}
\label{sec:modelStructure}

We view the programming paradigm of SCJ separately from its realisation in Java. We capture this paradigm, abstracting away from most of the details of its Java implementation. Our modelling approach is agnostic of Java.

We model the state and behaviour of application objects in the program and the use of suspension. We also capture exceptions, but not the Java exception handling mechanism. We only capture exceptions where they indicate a misuse of the paradigm. Specifically we capture exceptions when: a thread is interrupted, a thread attempts to use suspension without holding the lock, a thread attempts to lock an object with a priority lower than the thread's, a method receives an inappropriate argument, or a mission attempts to register a schedulable that is already registered to another or the same mission.

Our models consist of two parallel components, following the approach in~\cite{zeyda_etal_2013}. The framework component captures the behaviour of the library supporting the SCJ API and is reused for all programs. The application component captures the specific behaviour of a particular program. Each framework process has a counterpart application process. The complete specification of the framework model~\cite{SCJLevl2FW} comprises approximately 3700 lines of \Circus{} over 11 processes.

Table~\ref{circusOperators} summarises the \Circus{} action operators that we use in this paper. Most of them are familiar to users of CSP. We describe them to support the discussion of our model; a comprehensive account of \Circus{} is in~\cite{circus_woodcock_cavalcanti_2002}. We note that \Circus{} processes can also be combined using most CSP operators.

\begin{table}[t]
\centering
  \begin{tabular}{ l | l | l  }
    \hline
Action			& Syntax 								& Description \\ \hline \hline 
Skip			& $\Skip$ 								& A simple operator that terminates \\ \hline
Simple Prefix 	& $c \then A$ 							& Simple synchronisation with no data\\ \hline
Input Prefix 	& $c?x \then A$						 	& Synchronisation with a value bound to $x$ \\ \hline
Output Prefix	& $c!x \then A$ 						& Synchronisation outputting the value of $x$\\ \hline
Parameter Prefix			& $c.x \then A$ 						& Synchronisation with some data $x$ \\ \hline
Sequence		& $A \circseq B$  						& Executes $A$ then $B$ in sequence\\ \hline
External Choice & $A \extchoice B$ 						& Offers a choice between two actions $A$ and $B$\\ \hline
Interrupt		& $ A \circinterrupt c \then \Skip$		 & 	Executes $A$ unless $c$ occurs,  which terminates $A$\\ \hline
Recursion		& $\circmu X \circspot A \circseq X $ & A process $X$ that executes $A$ then $X$\\ \hline
Wait				& $\circwait t$	& Waits for $t$ time units and then terminates \\ \hline
Chaos			& $\Chaos$		& The action that immediately diverges \\ \hline \hline 
  \end{tabular}  
\caption{Summary of \Circus{} operators  \label{circusOperators}}
\end{table}

We describe our models in Sect.~\ref{sec:modelOverview} and present our approach in more detail in Sect.~\ref{sec:modelExample} using the mission models as an example. Finally, in Sect.~\ref{sec:SaS}, we discuss how we model synchronisation and suspension.

\subsection{Model Overview}
\label{sec:modelOverview}

Each SCJ library class and application object is represented by a \Circus{} process. Each process retains the name of the class it models, suffixed with `$FW$' for framework processes or `$App$' for application processes. Methods are represented by an action in the relevant process. Method calls and returns are represented by (usually pairs of) events; this allows method calls between processes.

\begin{figure}[t]
\centering
\includegraphics[width=\textwidth]{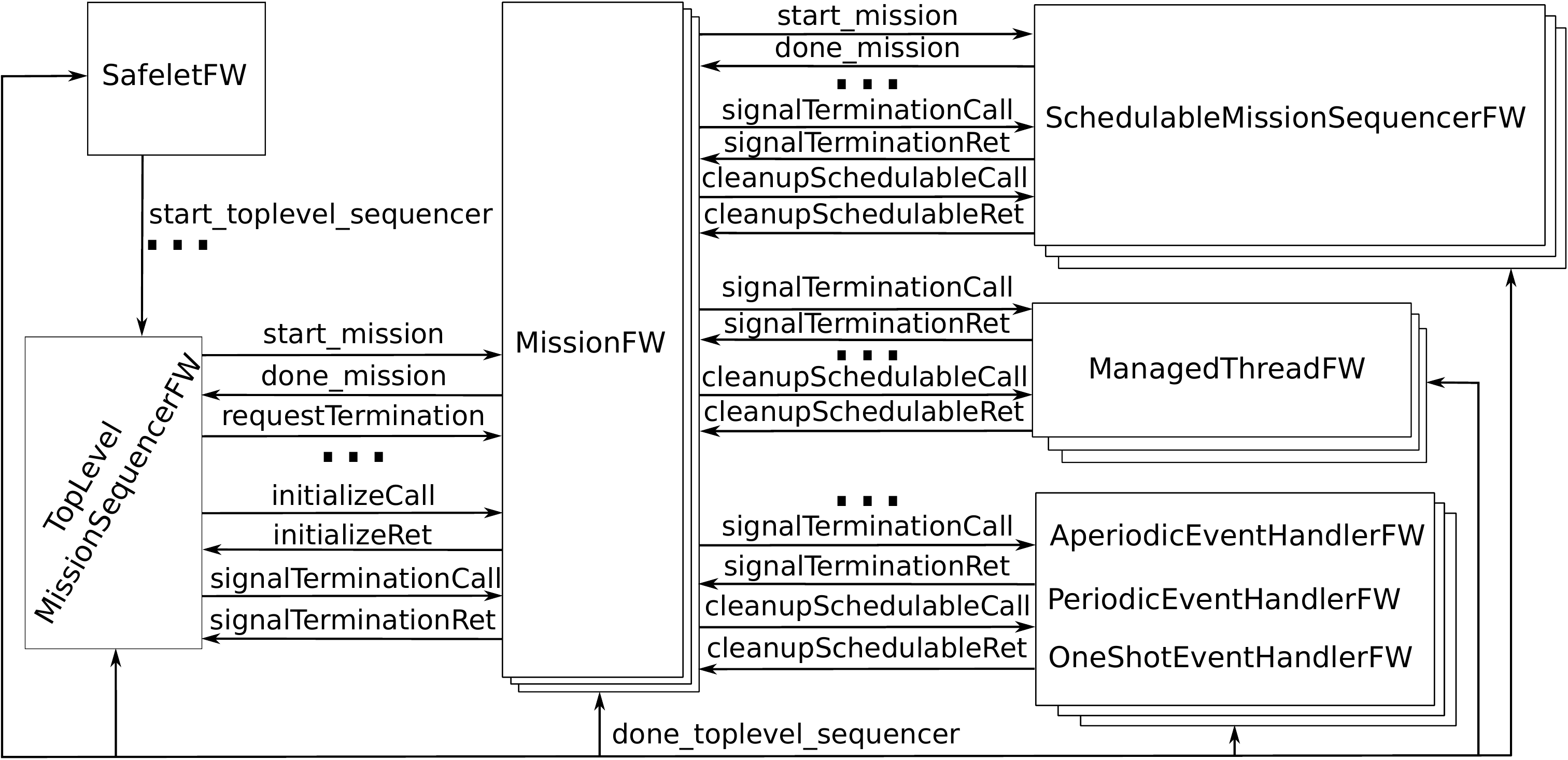}
\caption{Level 2 Model Structure \label{fig:Level2Structure}}
\end{figure}

Figure~\ref{fig:Level2Structure} shows the framework processes in our model and the channels that they use to communicate. The channels with underscores in their names are control signals (for example, $start\_mission$) and those in camel case represent method calls (for example, $initializeCall$ and $initializeRet$).
Some of the channels have been omitted for brevity, indicated by three dots. The layering indicates potentially multiple instances in one model. Each of these framework processes communicate with an application process; these are not shown in Fig.~\ref{fig:Level2Structure}.

When a framework process encounters application-specific behaviour, it signals its application counterpart to take control and perform the behaviour. Control is returned to the framework with another signal. These signals are call-return event pairs that retain the method name, suffixed with `$Call$', for the event modelling the method call, or `$Ret$', for the event modelling its return.

Each application process is assigned a unique identifier, allowing framework processes to communicate with their application counterparts. An exception is the $SafeletFW$ process, which only has one instance because there is only one safelet in an SCJ program. If a program class has multiple instances in the program, then each instance has its own \Circus{} process identifier. 

Our model uses \textsf{\textit{Oh}}\Circus{} classes to capture non-reactive behaviour, such as methods that are purely data operations. \textsf{\textit{Oh}}\Circus{} classes are similar to Java classes: they may hold variables, specify constructors, make use of inheritance, and must be instantiated before use. Specifically, data operations are captured in  methods, which may be called from processes. In contrast to \Circus{} processes, \textsf{\textit{Oh}}\Circus{} classes can be related by inheritance.

Instead of simply adding Level~2 features to the Level~1 model~\cite{zeyda_etal_2013}, we also capture Level~1 features not found in the previous model. Namely, we consider that a period or deadline may be overrun and capture exceptions and synchronisation. While Level~1 programs may not use suspension, they are allowed to use synchronisation. In addition, in contrast to the Level~1 model, we provide separate framework processes for each of the three kinds of event handlers, each encapsulating their particular release pattern. This simplifies the application models considerably and lessens the burden on translation. Further, as already mentioned, our model raises an exception if a schedulable is registered twice. 

\subsubsection{Safelet}

The framework process $SafeletFW$ handles the operations of the safelet. $SafeletFW$ gets the identifier of the top-level mission sequencer from its application counterpart and starts it. Additionally it raises an exception if the program attempts to register a schedulable that is already registered. This is the process that defines the main execution flow of the program. 

\subsubsection{Mission Sequencers}
Two framework processes model mission sequencers. The $TopLevelMissionSequencerFW$ process models the top-level mission sequencer and the $SchedulableMissionSequencerFW$ models a mission sequencer used as a schedulable. This simplifies both processes because they each only have to be involved in events relevant to their context.

Both flavours of mission sequencer fetch the identifier of the next mission from their application counterpart and start that mission. However, $SafeletFW$ starts $TopLevelMissionSequencerFW$, which signals to the entire model when it is terminating, to indicate that the program is done. 

$SchedulableMissionSequencerFW$ is started by a mission and signals to that mission once terminated. Since it is a schedulable, it must respond to termination requests from either its controlling mission or the mission it is executing. 

\begin{figure}[t]
\begin{circusaction}
InitializePhase \circdef \\
\quad \circblockopen
  initializeCall~.~FlatBufferMissionMID \then \\
	register~!~ReaderSID~!~FlatBufferMissionMID \then   \\
	register~!~WriterSID~!~FlatBufferMissionMID \then   \\
  initializeRet~.~FlatBufferMissionMID \then \\
  \Skip
\circblockclose
\end{circusaction}
\caption{The $FlatBufferMissionApp$'s $InitializePhase$ action \label{fig:initPhaseApp}}
\end{figure}

\subsubsection{Mission}

The $MissionFW$ process is started by a mission sequencer process. It then allows its application counterpart to register schedulables. It starts each schedulable and deals with their termination and clean\-up. If requested, it terminates itself and its active schedulables, and signals to its controlling mission sequencer that it has done. In Sect.~\ref{sec:modelExample} we describe the $MissionFW$ process in more detail and present the model of one of its actions.

\subsubsection{Schedulables}

Schedulables are modelled by $PeriodicEventHandlerFW$, for periodic event handlers; $AperiodicEventHandlerFW$, for aperiodic event handlers; $OneShotEventHandlerFW$, for one-shot event handlers; $ManagedThreadFW$, for managed threads; and $SchedulableMissionSequencerFW$, for mission sequencers used as schedulables. Each is started by a mission, performs its behaviour, accepts termination requests from its mission, and cleans up after it terminates. 

Each event handler has actions that control its specific release pattern. Event handlers may have deadlines associated with them, and periodic event handlers have an associated period. Our models consider that periods may be overrun and deadlines may be missed, and captures the response if this happens. This allows our models to be used to check if, for example, an event handler may overrun its deadline. Managed threads are simpler and begin their release as soon as they are started. Mission sequencers used as schedulables are described above.

\subsection{Mission Example}
\label{sec:modelExample}

The \Circus{} model of a mission is ideal to illustrate our modelling approach. The FlatBuffer application in Sect.~\ref{sec:exampleApp} contains one mission, \texttt{FlatBufferMission}, which we model using three components described next. 

As previously indicated, like every mission, an instance of $MissionFW$ represents the behaviour of the mission prescribed by the SCJ paradigm. It is outlined above. The non-reactive application-specific behaviour is captured in the \textsf{\textit{Oh}}\Circus{} class $FlatBufferMissionClass$. It contains the $buffer$ variable, corresponding to the buffer field of the \texttt{FlatBufferMission}, and the \texttt{bufferEmpty()} method, because it is purely a data operation without any reactive behaviour.

\begin{figure}[t]
\begin{circusaction}
Register ~ \circdef ~ \\
\quad  register~?~s~!~mission \then \\
\quad \circblockopen
	\circblockopen
		checkSchedulable~.~mission~?~check\prefixcolon(check = \true) \then\\
		\lschexpract AddSchedulable \rschexpract \\
	\circblockclose\\ 
	\extchoice \\
	\circblockopen
		checkSchedulable~.~mission~?~check\prefixcolon(check = \false) \then\\	
		throw.illegalStateException \then \\
  		\Chaos\\
  	\circblockclose
\circblockclose
\end{circusaction}
\caption{The $MissionFW$'s $Register$ action \label{fig:registerFW}}
\end{figure}

The $FlatBufferMissionApp$ process captures the reactive application-specific behaviour of the mission. It has actions modelling the behaviour of the API methods \texttt{initialize()} and \texttt{cleanup()} and actions modelling the application-defined methods: $writeSyncMeth$, $readSyncMeth$, and $bufferEmptyMeth$. It stores a reference to an instance of $FlatBufferMissionClass$, which contains the method \texttt{bufferEmpty()}.
The $bufferEmptyMeth$ action wraps this method, so that it can be called by other processes. 

Channels on which the instance of $MissionFW$ and $FlatBufferMissionApp$ communicate are parametrised by the mission identifier $FlatBufferMission$; this ensures that the $FlatBufferMissionApp$ communicates with the right framework process. The $FlatBufferMissionApp$ instantiates and communicates with the $FlatBufferMissionClass$ to call its \texttt{bufferEmpty()} method.

In an SCJ program, the \texttt{Mission}'s \texttt{initialize()} method is overridden to register the schedulables that this particular mission manages. In Fig.~\ref{fig:initPhaseApp} we show the $InitializePhase$ action of the $FlatBufferMissionApp$ process, which models the \texttt{initialize()} method in \texttt{FlatBufferMission}. The events $initializeCall$ and $initializeRet$ model the call to and return from \texttt{initialize()}. 

The registration of a schedulable is modelled by the event $register.s.m$, where $m$ is the identifier of the mission registering the schedulable and $s$ is the identifier of the schedulable being registered. The order of registration shown in Fig.~\ref{fig:initPhaseApp} corresponds to the order in the program. After registration, all registered schedulables are started simultaneously.

In $MissionFW$, $initializeCall$ triggers the $Register$ action (Fig.~\ref{fig:registerFW}), which accepts a $register$ event, with any schedulable identifier as long as the mission identifier is the same as this mission's. The $checkSchedulable$ event indicates, via the variable $check$, if the schedulable may be registered.

If $check$ is $\true$, then $Register$ can add the schedulable. If $check$ is $\false$, then the schedulable is already registered and we use the $throw$ channel to model an exception being thrown and then diverge ($\Chaos$). This allows the detection of an attempt to register a schedulable more than once.

\subsection{Synchronisation and Suspension}
\label{sec:SaS}

The synchronisation model of SCJ constrains that of Java. First, SCJ programs cannot use \texttt{synchronized} blocks, only \texttt{synchronized} methods. Second, threads queue for a lock in order of eligibility. In SCJ, the most eligible thread is the thread at the highest priority level that has been waiting for the longest time. We model this using the type $PriorityQueue$, which is a total function from $PriorityLevel$ to injective sequences of $ThreadID$. $PriorityLevel$ is a free type containing the priorities available to the system and $ThreadID$ is the set of thread identifiers.

Our models use extra processes to control synchronisation and suspension. In SCJ, each schedulable is executed by a thread. In our model, schedulables that call a synchronised method are associated with an instance of the $ThreadFW$ process. $ThreadFW$ holds the thread identifier and keeps track of its priority and interrupted status. Overall, the framework model of a schedulable that calls a synchronised method is the parallel composition of its associated $ThreadFW$ process with the appropriate framework process, which depends on the type of schedulable (event handler, managed thread, and so on). 

Additionally, each object used as a lock is associated with an instance of the $ObjectFW$ process, which stores the threads waiting on this object and controls the threads trying to lock this object. In the FlatBuffer, the mission is used as a lock, so it has an associated instance of $ObjectFW$. Again, the overall framework model of each object that represents a paradigm component and is used as a lock is its framework process in parallel with an instance of $ObjectFW$. Non-paradigm objects used as locks are modelled in the framework by just an instance of $ObjectFW$.

The FlatBuffer program uses synchronisation and suspension to control access to the buffer in its mission. The synchronised \texttt{read()} method suspends the calling thread (by calling \texttt{wait()}) if the buffer is empty. This is wrapped in a loop that checks if the buffer is empty, to deal with  spurious wake ups.
 
The \texttt{FlatBufferMission}'s \texttt{read()} method is modelled by the $readSyncMeth$ action in the $FlatBufferMissionApp$ process (Fig.~\ref{fig:readSyncMeth}), which shows the pattern we use for modelling all synchronised methods. The action begins and ends with the familiar call-return event pair, $readCall$ and $readRet$, which correspond to the call to and return from the method. In this case, however, because this is a synchronised method, these events take an extra parameter $thread$, which is the identifier of the thread that is calling the method. 

\begin{figure}[t]
\begin{circusaction}
readSyncMeth \circdef \circvar ret : \num \circspot \\
\quad \circblockopen
readCall~.~FlatBufferMissionMID~?~caller~?~thread \then \\
\circblockopen
startSyncMeth~.~FlatBufferMissionOID~.~thread \then \\
lockAcquired~.~FlatBufferMissionOID~.~thread \then \\
\circblockopen
\circblockopen 
\circmu X \circspot \circblockopen \circvar loopVar : \boolean \circspot loopVar :=~bufferEmpty()\circseq \\ 
\circif ~ (loopVar = \true) ~ \circthen \\
 \quad \circblockopen waitCall~.~FlatBufferMissionOID~.~thread \then \\ 
 waitRet~.~FlatBufferMissionOID~.~thread \then \\ 
 \Skip \circblockclose \circseq X \\
 \circelse ~ (loopVar = \false) \circthen \Skip \\ 
 \circfi \circblockclose \circblockclose \circseq \\
\circvar out : \num \circspot out := this~.~buffer \circseq \\
this~.~buffer :=0 \circseq \\
notify~.~FlatBufferMissionOID~!~thread \then  \\
  ret := out
        \circblockclose
 \circseq  \\
endSyncMeth~.~FlatBufferMissionOID~.~thread \then  \\
readRet~.~FlatBufferMissionMID~.~caller~.~thread~!~ret \then \Skip
\circblockclose
\circblockclose
\end{circusaction}
\caption{The $FlatBufferMission$ proceess's $readSyncMeth$ action \label{fig:readSyncMeth}}
\end{figure}

The $ObjectFW$ process associated with the $FlatBufferMissionApp$ process controls the synchronisation and suspension behaviour using the $startSyncMeth$, $lockAcquired$, and $endSyncMeth$ events. The $startSyncMeth$ event models the beginning of a synchronized method and triggers the $ObjectFW$ process to request a lock on this object by the thread calling this action.

Because the lock may already be held by another thread, the $readSyncMeth$ action waits for the $lockAcquired$ event (from the $ObjectFW$ process) to signal that it has the lock and can proceed. After the body of the method, the $endSync$ event signals that the synchronised method is complete, to trigger $ObjectFW$ to release the lock on the mission currently held by the calling thread. We note that SCJ does not support Java's \texttt{ReentrantLock}, however, SCJ does support reentrant locking  by allowing synchronised methods to call other synchronised methods in the same object. The $ObjectFW$ process provides this behaviour; to unlock the object, after the first $lockAquired$ event, each subsequent $startSyncMeth$ event (which must be from the same thread) must be matched by a $endSyncMeth$ event from the locking thread. 

We model the call to \texttt{wait()} using the call-return event pair $waitCall$ and $waitRet$. These events take the identifier of the associated $ObjectFW$ instance ($FlatBufferMissionOID$, in Fig.~\ref{fig:readSyncMeth}) and the identifier of the $thread$ calling this action. The instance of $ObjectFW$ associated with the mission adds $thread$ to its queue of waiting threads. The process calling $waitCall$ waits for $waitRet$ to communicate its identifier.

We model the call to \texttt{notify()} using the event $notify$. Like $waitCall$ and $waitRet$, this event also takes the identifier of the associated $ObjectFW$ process and the identifier of the $thread$ calling this action. The $notify$ event triggers the $ObjectFW$ process to resume the most eligible thread. If there are no waiting threads, then $ObjectFW$ allows the call to $notify$, but does nothing. To resume a thread, $ObjectFW$ calls $waitRet$ with the identifier of the thread to be resumed. SCJ Level~2 can also use \texttt{notifyAll()}, which resumes all the waiting threads. We model a call to \texttt{notifyAll()} with the event $notifyAll$. It triggers the $ObjectFW$ to call $waitRet$ with the identifier of each waiting thread in eligibility order.

The complete \Circus{} models of the framework processes can be found in~\cite{SCJLevl2FW}, and the application processes of the FlatBuffer in~\cite{hiJaCCaseStudies}. In the next section, we discuss the validation and application of our models.

\section{Initial Evaluation}
\label{sec:applications}

Our \Circus{} model is written to closely correspond with the SCJ API. We have frozen development of our model at version 0.100 of the SCJ language specification. One of the authors is a member of the SCJ Expert Group, which helped in clarifying ambiguities in the language specification.

Our model of Level~2 is based on the \Circus{} model of Level~1 presented in~\cite{zeyda_etal_2013}, which has been validated against the SCJ language specification. Our model adds the features of Level~2 and updates the model to reflect recent changes in the language specification.

Our modelling effort has influenced the development of SCJ. In~\cite{SCJL2inPractice_submitted}, which is under review, we present a model of the SCJ termination protocol and a proposed simplified termination protocol. The comparison of these models shows that our proposed protocol reduces the number of states in the system. This simplified protocol is useful for improving programmer understanding and further modelling efforts. Our simplified termination protocol was adopted by the SCJ expert group from version 0.96.

We have, by hand, translated 10 SCJ programs to \Circus{} using our approach; the examples are summarised in Table~\ref{tab:handTrans}. The programs are constructed to cover the features of SCJ. They range from simple tests of SCJ's features, such as different release patterns or synchronisation and suspension, to more complex programs that use nested mission sequencers to provide concurrent missions. 

Further, we have developed a prototype tool\footnote{\TightRope~can be found at~\url{www.cs.york.ac.uk/circus/hijac/tools.html}.} to automatically generate the \Circus{} application models of a given SCJ application, called \TightRope. We have used this prototype to produce the application models of the FlatBuffer application presented in this paper and a more complex example, both summarised in Table~\ref{tab:toolTrans}. The 10 hand-translated examples, and more realistic programs, will be considered for automatic translation as \TightRope~matures.

\begin{table}[t]
\centering
  \begin{tabular}{  c | p{7.5cm} | c  }  
  Name & Description & \textnumero~ Classes  \\ \hline \hline
   Mission1 & A single mission with periodic event handler that releases an aperiodic event handler & 5  \\ \hline
   Mission2 		& A single mission with a managed thread and a one-shot event handler & 5  \\ \hline
   ThreeOneShots 	& A single mission with three one-shot event handlers & 6   \\ \hline
   ThreeThreads  	& A single mission with three managed threads & 6  \\\hline
   SequentialMissions & Two sequential missions, each with two managed threads & 8  \\ \hline
   NestedSequencer1 & A single mission with a single nested mission sequencer & 7 \\ \hline
   NestedSequencer2 & A mission, with three nested mission sequencers. Each has one mission controlling a periodic event handler & 14 \\ \hline
   NestedSequencer3 & A mission, with a nested mission sequencer that has two sequential nested missions, each with a managed thread. & 8 \\ \hline
   NestedSequencer4 & A complicated example using two levels of nesting. It contains 4 missions and 3 managed threads & 12  \\ \hline
  NestedSequencer5 & Extends NestedSequencer4, combines complex nesting, all schedulable types, and sequential missions & 12  \\ \hline
\end{tabular}
\caption{Summary of SCJ programs translated by hand \label{tab:handTrans}}
\end{table}

\TightRope~is a small Java program that compiles an SCJ application and explores the resulting abstract syntax trees to extract the information required for the translation. \TightRope~generates the \Circus{} processes, \textsf{\textit{Oh}}\Circus{} classes, and \Circus{} channels required to model the application-specific behaviour of the input program. These are combined with the existing fixed models of the framework previously described, to form a specification of the whole program.

To facilitate model checking and animation using FDR3~\cite{fdr3}, we have translated our models of the framework and of full programs into \texttt{CSPm}. This translation has been optimised so that FDR3 can check specifications of even complex programs in an acceptable amount of time. We have proved that the \texttt{CSPm} version of the framework model is deadlock- and divergence-free, which lends extra validation to the framework. We have also proved that the models of the full programs that we translated do not deadlock or diverge. 

Using the version of the CSP animator ProBE that is included in FDR3, we have animated the \texttt{CSPm} versions of the framework model and compared their behaviour with that prescribed in the SCJ language specification. This gives us confidence that the models capture the behaviour of the SCJ API. We have also used ProBE to examine the behaviour of these full models, to compare them to the running programs. We have compared the execution of our example SCJ Level~2 programs, using the IceLab~\cite{IceLabSCJ} implementation, to animations of our models of these programs. These comparisons examined the behaviour and output from the executing programs with the corresponding events in the animated model to ensure that they have the same behaviour. 

Future work in the analysis of our models includes extending the checks we make to cover more SCJ-specific criteria. We intend to check that the program does not attempt to register its top-level mission sequencer or throw any of the exceptions that we model. Because we model exceptions using an event followed by divergence, they are flagged by a divergence-freedom check. However, the counter examples provided by a specific check would be more useful during SCJ development. These SCJ-specific checks will be standardised for easy reuse.

In summary, because our framework model captures the behaviour of the SCJ paradigm separately from the program-specific behaviour, we can reason about it in isolation. We have used FDR3 to prove that the framework model does not deadlock or diverge. Models of full SCJ Level~2 programs can be model checked and animated in FDR3. Our formal semantics of the Level~2 paradigm enables further areas of study for SCJ Level~2, such as theorem proving. 

\begin{table}[t]
\centering
  \begin{tabular}{  c | p{6cm} | c | c  }
  	Name & Description & \textnumero~ Classes & Translation Time \\ \hline \hline
	 FlatBuffer	& Small program using managed threads and synchronisation & 6 & ~1.2 seconds \\ \hline
  	Aircraft & Program using a schedulable mission sequencer to represent phases of aircraft flight &25 &  ~2.3 seconds  \\ \hline   
  \end{tabular}
\caption{Summary of SCJ programs translated by \TightRope \label{tab:toolTrans} }
\end{table}

\section{Related Work}
\label{sec:related}

This is the first work supporting verification for SCJ Level~2 programs. K-Java~\cite{Bogdanas2015-de} models a subset of SE Java 1.4 and produces executable specifications for model checking. However, SCJ programs have features not included in SE Java. The authors of~\cite{Thomsen2015-dg} present a technique for translating SCJ programs into timed automata models. However, their technique appears to only be aimed at Levels~0~and~1. Further, neither of these techniques provide support for top-down refinement of SCJ Level~2 programs or refinement-based reasoning.  

RSJ~\cite{kalibera_etal_2010} is a adaptation of the Java PathFinder~\cite{havelund_pressburger_2000} that explores all possible schedulings of the threads within an SCJ program to check for scheduling-dependent errors. It, however, does not cater for SCJ Level~2 programs.

Older versions of the SCJ specification define annotations for specifying compliance level, behavioural, or memory restrictions. Previous approaches to ensuring the safety of SCJ programs have used these annotations to provide run-time checks~\cite{tang_static_2010} or to specify checkable program constraints~\cite{Haddad_2010}. However, the memory annotations have been moved to an appendix of the standard as they were judged not ready for standardisation.

Our modelling approach is similar to that of~\cite{zeyda_etal_2013} in capturing the paradigm of SCJ Level~1. The underlying structure of programs written in Level~2 and Level~1 is the same, however, Level~2 allows much more complicated program hierarchies and provides more complicated features (such as suspension). 

\section{Conclusion}
\label{sec:conclusion}

We have presented the first formal semantics of SCJ Level~2, using the \Circus{} family of specification languages. It is an essential ingredient to enable customised top-down development of SCJ Level~2 programs that are correct by construction. Our models provide this development process with a target for SCJ Level~2.

The features \Circus{} provides make it a good fit for modelling object-orientated languages, such as SCJ. A \Circus{} process provides similar encapsulation to classes and the language can capture variables and methods. This means that our models correspond very closely to the programs they model. 

We have validated our model of the SCJ API and Level~2 programs by translating them into \texttt{CSPm} and model checking it using FDR3 to show that it does not deadlock or diverge. Our prototype tool, called \TightRope, has produced \Circus{} models of SCJ applications. Work is ongoing to update the tool, so that it can generate the models for all of our example applications. 

In addition to the further areas of study that our work enables, future work includes the formalisation of the translation strategy that we use to derive the application models from the SCJ programs. The translation strategy also needs to be evaluated on more applications to further test our modelling approach.

\section*{Acknowledgements}

This research reported in this paper is funded by the UK EPSRC under grant EP/H017461/1. No new primary data was produced during this work. One of the authors is a member of the SCJ Expert Group; we would like to thank the other members of the Expert Group. We would also like to thank Frank Zeyda, Alan Burns, and Thomas Gibson-Robinson for their very helpful suggestions.

\bibliographystyle{splncs03}
\bibliography{overviewPaper}

\end{document}